\documentclass{article}

\usepackage[utf8]{inputenc}
\usepackage{graphicx}

\usepackage{authblk}
\usepackage{tikz}
\usepackage[normalem]{ulem}

\usepackage{array,makecell}

\usepackage{etoolbox}

\usepackage[hyphens]{url}

\usepackage{todonotes}

\usepackage{fancyhdr}

\usepackage{xcolor}
\usepackage{tcolorbox}
\usepackage{titlesec}
\usepackage{adjustbox}
\usepackage{rotating}
\usepackage{multirow}

\definecolor{chaptextbg}{RGB}{220, 220, 220}
\definecolor{chaptext}{RGB}{150, 150, 150}

\newcommand{\printbibliography}[1]{%
  \bibliographystyle{alpha}
  \bibliography{#1}
}

\title{Energy-Aware Workflow Execution: An Overview of Techniques for Saving Energy and Emissions in Scientific Compute Clusters}

\date{}

\begin{document}

\author[1]{Lauritz Thamsen}
\author[1]{Yehia Elkhatib}
\author[1]{Paul Harvey}
\author[1]{Syed Waqar Nabi}
\author[1]{Jeremy Singer}
\author[1]{Wim Vanderbauwhede}

\affil[1]{University of Glasgow, School of Computing Science, Glasgow, United Kingdom}

\maketitle

\begin{abstract}
Scientific research in many fields routinely requires the analysis of large datasets, and scientists often employ workflow systems to leverage clusters of computers for their data analysis.
However, due to their size and scale, these workflow applications can have a considerable environmental footprint in terms of compute resource use, energy consumption, and carbon emissions. Mitigating this is critical in light of climate change and the urgent need to reduce carbon emissions. 

In this chapter, we exemplify the problem by estimating the carbon footprint of three real-world scientific workflows from different scientific domains.
We then describe techniques for reducing the energy consumption and, thereby, carbon footprint of individual workflow tasks and entire workflow applications, such as using energy-efficient heterogeneous architectures, generating optimised code, scaling processor voltages and frequencies, consolidating workloads on shared cluster nodes, and scheduling workloads for optimised energy efficiency.
\end{abstract}

\section{Introduction} 

The carbon footprint of ICT is rising despite the urgent need to decarbonise society and to stay within planetary tolerance levels~\cite{belkhir2018assessing, masanet2020recalibrating,Freitag2021ClimateImpactICT}.
One factor contributing to the rise is the growing amount of data collected, stored, and processed – as it is increasingly done in many sciences.

Scientific workflow management systems (SWMSs) enable users to leverage clusters of computers for processing large datasets.
This is becoming more and more important in many fields, including remote sensing, astronomy, and bioinformatics.
However, scientific workflow applications can be long-running, large-scale, and resource-intensive.
Hence, executing workflows can consume substantial amounts of energy as well as require substantial compute infrastructure, both of which cause greenhouse gas emissions.

There are various techniques to lower the energy consumption and carbon footprint of such applications, including generating energy-efficient code for workflow tasks and optimising for energy efficiency while scheduling workflow applications.
Yet, it is difficult for practitioners and researchers to gauge which options are applicable, will make a meaningful difference, and can be practically implemented without requiring a prohibitive effort.
To shed some light on these issues, this chapter provides an overview of techniques for executing workflows so that they consume less energy and, in turn, cause fewer emissions. Our focus is, therefore, on the operational carbon emissions from the energy consumption of running workflows on clusters, rather than embodied carbon in cluster hardware or the carbon costs of development and deployment.

In the following, we first summarise our working model of scientific workflows and relevant sustainable computing background (Section~\ref{sec:background}). Then, we analyse the problem by demonstrating how one can calculate a rough estimate of the carbon footprint of real-world scientific workflows (Section~\ref{sec:problem_analysis}). Subsequently, we present general techniques and recent works for improving the energy efficiency of individual workflow tasks (Section~\ref{sec:individual_task_techniques}) and entire workflow applications (Section~\ref{sec:entire_workflow_techniques}). Finally, before the conclusion, we discuss the impact of available compute infrastructure and energy sources as well as economical implications (Section~\ref{sec:discussion}).

\section{Background}
\label{sec:background}

This section presents the working model of scientific workflows assumed in this chapter and key concepts and considerations related to the sustainability of computing.

\subsection{Scientific Workflows}
\label{sec:workflow_model}

Scientific workflow management systems (SWMSs), such as Nextflow, Pegasus, and Snakemake, allow users to construct larger workflow applications from individual data processing tasks. 
These tasks can be arbitrary programs and are generally treated as black boxes by SWMSs.
Workflow applications then take the form of directed acyclic graphs, connecting individual tasks to other tasks via ``channels". 
That is, the output of one task becomes the input of one or more subsequent tasks. 

\begin{figure}[h]
  \centering
  \includegraphics[width=0.8\textwidth]{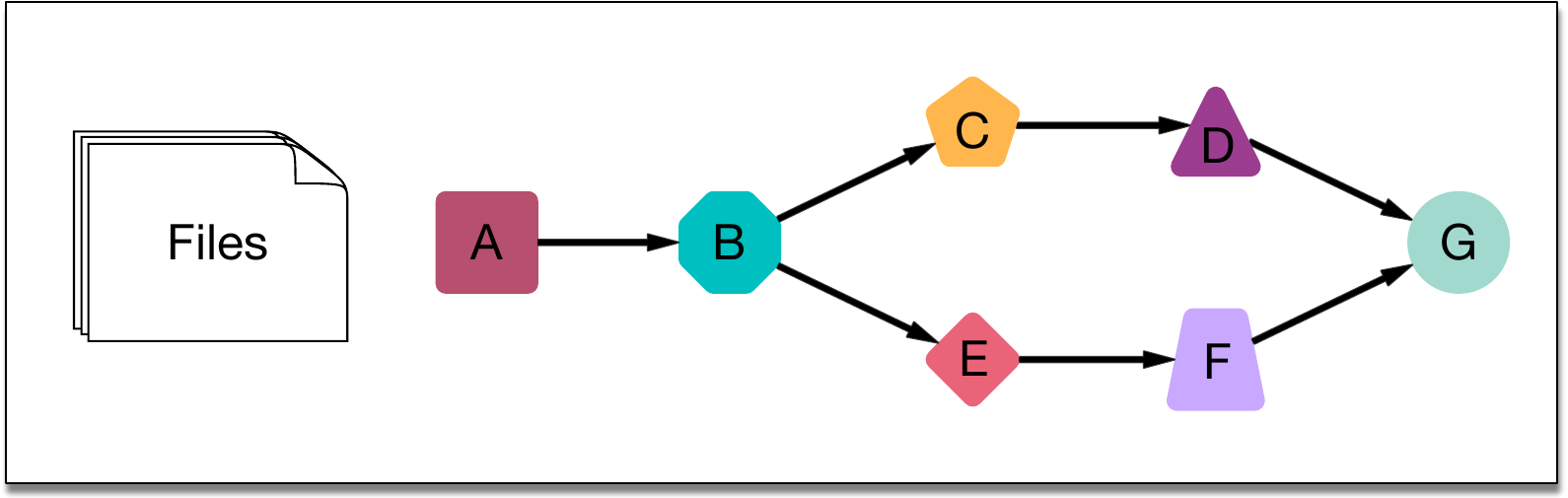}
  \caption{A workflow application that connects seven data processing tasks through which incoming data passes successively.}
  \label{fig:workflow_application}
\end{figure}

An example of a workflow application graph that connects multiple tasks in this way is shown in Figure~\ref{fig:workflow_application}. It has an initial task A that reads file inputs and passes its results to task B via a channel. Here, the workflow graph forks into two paths and the outputs of task B are processed by both tasks C and E. After two further processing steps, D and F, the data is merged again in the workflow's final task, G.

Workflow applications are typically executed per input, or set of related inputs, allowing coarse-grained data parallelism at the application level, with multiple tasks running simultaneously if there are multiple independent inputs to process using the same workflow application.
In addition, if a task in a workflow application graph outputs data to multiple subsequent tasks so that the workflow graph forks, multiple tasks can process data in parallel even for just one original input to the workflow application. Similarly, some workflows start by executing multiple different tasks, which can run in parallel, before their outputs are eventually required in a task that effectively joins and synchronises multiple workflow paths.

Many workflow applications contain a large number of tasks, include multiple paths that can run in parallel, and are executed on many independent inputs.
Therefore, workflow applications are commonly executed on clusters, which are groups of compute nodes that are locally connected and collectively managed.
For this, SWMSs typically submit the tasks of a workflow that are ready to run to a cluster resource manager such as Slurm or Kubernetes.
The cluster resource manager then allocates tasks to nodes, respecting user-specified resource requirements, such as the number of cores and memory required.
Meanwhile, the channels between tasks are often realised by writing to and reading from persistent cluster storage systems such as Ceph, HDFS, or networked storage systems such as NFS.
This enables a simple execution model, flexibility in scheduling, and the ability to restart failed tasks from intermediate results stored safely on disk.

While SWMSs simply run tasks with the inputs and parameters that workflow applications define, it is sometimes beneficial for researchers and developers to open up these black boxes.
For example, workflow performance can sometimes be improved significantly by optimising the efficiency of key tasks for the available hardware, for example by generating code specifically for the available CPUs or accelerators.
Often, some code blocks impact a task's overall runtime more than others, making them a good starting point for optimisations.

\subsection{Sustainable Computing Concepts and Considerations}
\label{sec:computing_footprint}

In this section, we summarise key concepts for sustainable computing and illustrate the energy consumption of major computer system components.

\paragraph{Static and dynamic power consumption.} 
Computer system components draw power when turned on but idle. This is referred to as \emph{static} power consumption.
Meanwhile, there is typically also a load-dependent, \emph{dynamic} power consumption, correlated to resource utilisation. The total power consumption is then the sum of the static and dynamic parts. The static energy consumption of servers can be substantial, such as up to half of its peak power consumption under maximum load; that is, significant power is wasted on resources that do not perform any active work.

\paragraph{Energy-proportional computing.} Computing hardware is said to be energy proportional when its energy consumption per operation is independent of the current utilisation~\cite{Barroso_2019_DatacenterComputer}. This goal is in contrast to a high static power consumption, which will be more significant when the utilisation is low. Over the last two decades, server hardware has become more energy proportional.
Power-saving techniques such as Dynamic Voltage and Frequency Scaling (DVFS) can make processors even more energy-proportional. Another approach is to consolidate workloads onto fewer shared resources to achieve a higher and more consistent utilisation, 
 thereby improving energy proportionality. 

\paragraph{Power usage effectiveness.} Usually, there is not only the static and dynamic power consumption of compute, storage, and networking equipment, but also adjacent power consumption such as for cooling, lighting, and access control.
Reflecting this, \emph{Power Usage Effectiveness (PUE)}~\cite{Barroso_2019_DatacenterComputer} is a measure of how much energy in a data centre is required in addition to the energy consumed by the computing equipment. 
A PUE of 1.0 means that all the power entering a data center is used only by the computing equipment. 
Modern hyperscale data centres can have PUEs of around 1.1, while the estimated average for data centres globally in 2021 was 1.6.\footnote{\url{https://uptimeinstitute.com/resources/asset/2021-data-center-industry-survey}, accessed 2024-03-09}
However, it should be noted that the PUE of data centres is highly dependent on the exact measurement methodology and the workload. The best results tend to be achieved with constant loads of high utilisation. Similarly, PUE as a metric benefits from inefficiently operating IT equipment, as more power will then be used by it, and PUE does not reflect whole system trade-offs, such as the potential benefits of heat recovery.

\paragraph{Energy carbon intensity.} Depending on the energy mix, i.e. how much electricity is generated from fossil fuels, nuclear energy, and renewable sources, there are more or less emissions associated with the use of electricity. This is captured by the energy's \emph{carbon intensity}, which measures the $CO_{2}$-equivalent greenhouse gas emissions released for the energy consumed. It is usually expressed in grams of $CO_{2}e$ per kWh. It often varies between locations and over time in response to variable demand and fluctuations in renewable generation from solar and wind sources. 
For instance, in 2022, the average carbon intensity of the grid in the UK was 235 g$CO_{2}e$/kWh, with Germany at 473, France at 90, and Sweden at 24.\footnote{Aggregated data for the year 2022 as presented by \url{https://app.electricitymaps.com}}
Such yearly national aggregates are often used for carbon accounting, even though the average carbon intensity can fluctuate significantly from one hour to the next and also across regions within countries.
In the UK, for instance, the average carbon intensity has varied between below 100 and more than 300 g$CO_{2}e$/kWh in recent years, while it can fall below 10 for regions with abundant renewable wind power such as South Scotland.

\paragraph{Computer component energy consumption.} 
Computer system components such as processors, memory, disks as well as networking equipment consume energy, which in turn can be linked to greenhouse gas emissions.

\begin{itemize}
    \item \textbf{CPU}: Often a considerable share of the total power consumed by a computer goes to CPUs. For instance, an x86 processor can make up between 40\% and 66\% of a server's full power consumption~\cite{Barroso_2019_DatacenterComputer}, depending on node utilisation. Moreover, CPUs can have a high static power consumption, independent of utilisation, as well as a considerable dynamic power consumption, dependent on utilisation. For example, Intel's Xeon E5-2660 v3 was measured to consume between around 50 (idle) and 150 (peak) W at 2.5 GHz~\cite{Song_2025_HaswellPower}.
    \item \textbf{Memory}: Memory is often the second largest individual contributor to a computer system's energy consumption, but typically contributes only an eighth to a sixth %
    of a server's overall power draw~\cite{Barroso_2019_DatacenterComputer}. Memory is also usually less dependent on utilisation than CPUs are. The peak power consumption of a server's memory is often only twice as high as its idle state~\cite{Barroso_2019_DatacenterComputer}. Meanwhile, load-independent estimates for the power consumption of DDR3 and DDR4 memory are, for example, around 3 W per 8 GB.\footnote{\url{https://www.crucial.com/support/articles-faq-memory/how-much-power-does-memory-use}, accessed 2023-10-27}
    \item \textbf{Storage}:
    HDDs usually have even less of a dynamic power range, where the peak load can be just 30\% more than the idle load~\cite{Barroso_2019_DatacenterComputer}. The dynamic power consumption of other components can mean that HDDs are responsible for around 15\% of the overall power draw in idling systems, but only around 5\% at peak system utilisation. A 2016 US Data Centre Energy Usage Report\footnote{\url{https://eta.lbl.gov/publications/united-states-data-center-energy}, accessed 2023-10-27} estimates that in 2020 both HDDs and SSDs require not much more than 6 W per disk. However, since HDDs still commonly offer twice the capacity of SSDs, storing data on HDDs is often twice as energy-efficient. Nevertheless, we note that SSDs are typically more energy proportional, meaning their energy consumption more accurately reflects actual use~\cite{powerProportionalStorage}.
    \item \textbf{Networking}: Equipment like switches have a very low dynamic power consumption, even lower than disks, often using less than 20\% more energy at peak load~\cite{Barroso_2019_DatacenterComputer}. At the same time, in larger data centres, switches are typically used by many servers. This makes attributing the energy consumption of switches to specific applications not straightforward. Consequently, data centre application footprint methodologies such as Cloud Carbon Footprint\footnote{\url{https://www.cloudcarbonfootprint.org/docs/methodology}} simply exclude networking.
\end{itemize}

\paragraph{Full lifecycle emissions.} Aside of the \emph{operational emissions} from the energy used to power computing systems, the full lifecycle emissions also include \emph{embodied carbon emissions} and \emph{end-of-life emissions}:

\begin{itemize}
    \item \textbf{Embodied carbon emissions:}
The embodied carbon emissions stem mainly from material sourcing and manufacturing, and they can be substantial.
For example, the embodied carbon emissions of Intel and AMD processors can easily be comparable to operational emissions~\cite{Gupta_2021_ChasingCarbon}.
This assumes a carbon intensity during the use phase of the processors that is around the current global average.
However, with energy mostly from renewable sources or nuclear during the years of use, embodied carbon can even considerably outweigh operational emissions.
For instance, if the energy consumed is generated from wind, around 80\% of the overall emissions for x86 processors would be embodied~\cite{Gupta_2021_ChasingCarbon}. 
A full lifecycle assessment of the Dell R740 server\footnote{\url{https://www.delltechnologies.com/asset/en-us/products/servers/technical-support/Full_LCA_Dell_R740.pdf}, accessed 2023-10-15} similarly sees embodied and operational emissions on par, using current grid carbon intensities for the operational emissions from energy consumption and assuming four years of use.
Notably, almost 80\% of the embodied carbon of the Dell server are attributed to its SSDs. 
Generally, the embodied carbon rises linearly with the capacity of the SSD~\cite{tannu2023dirty}. 
It is therefore important to realise that wasting disk space in clusters is effectively causing emissions, because it forces the cluster operators to install more disk capacity. 

    \item \textbf{End-of-life emissions:}
Compared to operational and embodied carbon emissions, end-of-life emissions currently do not make up a major share of the total emissions. Emissions for end-of-life processing and materials recycling is often responsible for not more than 5\% of the full lifecycle footprint of computer systems~\cite{Gupta_2021_ChasingCarbon}.
Meanwhile, the Dell server lifecycle assessment even estimates that recycling reduces embodied emissions by around 1.8\%.
\end{itemize}

As the focus of this chapter is on the execution of workflow tasks and applications, we will concentrate on operational energy use and its emissions in the following.

\section{Estimating The Energy Consumption and Carbon Emissions of Scientific Workflows}
\label{sec:problem_analysis}

In this section, we exemplify the carbon footprint of scientific workflow applications by estimating their energy consumption and greenhouse gas emissions.
We first look at a geoscience workflow executed on a commodity cluster, then analyse an astronomy workflow executed on a major public cloud, and finally examine an individual bioinformatics workflow task executed on both commodity cluster and public cloud resources.

\subsection{Estimation Methodology}
\label{sec:footprint_estimation_methodology}

The operational emissions of a computing application can be estimated based on its resource usage.
For the resources allocated and used by an application, we can sum up the resources' static and dynamic power consumption to obtain a total power consumption \( P_{total} \) as in Equation~\ref{eq:power-total}.
The static power consumption \( P_{static} \) is independent of resource utilisation, and measures the idle power consumption of allocated resources regardless of usage.
If resources are shared by multiple applications, the static power consumption can be attributed proportionally to the share allocated to each application.
The dynamic power consumption \( P_{dynamic} \) is directly dependent on the resource utilisation of a specific application.
We resort to linear models for the dynamic power consumption of compute resources, interpolating linearly between static and peak power consumption, as shown in Equation~\ref{eq:power-dyn}.
Subsequently, the power consumption of an application \( P_{total} \) can be multiplied by the power usage effectiveness \( \textup{PUE} \) of the compute environment and the energy carbon intensity \( \textup{CI} \) of the electricity grid.
This gives us an estimate of the carbon-equivalent greenhouse gas emissions \( E \) in terms of Equation~\ref{eq:emissions}.

\begin{equation} \label{eq:power-total}
P_{total} =  P_{static} + P_{dynamic}
\end{equation}

\begin{equation} \label{eq:power-dyn}
P_{dynamic} =  (P_{peak} - P_{static}) \cdot Utilisation
\end{equation}

\begin{equation} \label{eq:emissions}
E = P_{total} \cdot \textup{PUE} \cdot \textup{CI}
\end{equation}

Estimating the operational emissions of software applications by multiplying power consumption with PUE and CI follows the Software Carbon Intensity (SCI) specification of the Green Software Foundation\footnote{\url{https://sci.greensoftware.foundation/}, Version 1.1.0} and is also implemented, for example, by the Cloud Carbon Footprint (CCF) methodology \footnote{\url{https://www.cloudcarbonfootprint.org/docs/methodology}} and the Green Algorithms (GA) project\footnote{\url{https://www.green-algorithms.org/}}~\cite{GreenAlgorithms}.
In addition, both CCF and GA use linear models to estimate the dynamic power consumption of compute resources based on resource utilisation (whereas the SCI leaves open how the power consumption is to be calculated).
Approximating power consumption of workflows in this manner should allow for reasonable ballpark estimates in comparison to established software power meters like Intel's RAPL.
For example, our research recently showed that such estimates have an error between 14.4\% and 47.9\% for a popular real-world Nextflow workflow~\cite{west2024ichnoscarbonfootprintestimator}, depending on the processing hardware and its configuration.

We directly use CCF methodology estimates for public cloud compute resources, and we specify all our sources of static and peak power measurements for processors in bare-metal clusters.
Furthermore, we use CCF coefficients for translating memory and disk usage into power consumption, as these estimates are general for data centres, so apply beyond specific public cloud resources.
For carbon intensity averages, we leverage data aggregated by ElectricityMaps\footnote{\url{https://app.electricitymaps.com}}.
Like CCF and GA, we use national annual average CI data as this is still more commonly used for carbon emissions accounting, even though finer-grained CI data better captures the fluctuating availability of low-carbon energy over time and across different locations.

\subsection{A Geoscience Workflow on a Commodity Cluster}

\paragraph{Workflow application.}
The first example is an Earth observation data processing workflow called FORCE~\cite{Frantz_2019_FORCE} that was implemented in Nextflow~\cite{lehmannFORCENextflowScalable2021}.
This implementation was evaluated with several scale-out experiments, running the workflow on 304 GB of satellite images of the island of Crete on up to 21 nodes on a cluster at TU Berlin.
We estimate the carbon footprint of the workflow's execution on 21 nodes, which ran for 315 minutes with an average CPU utilisation of 50\%.

\paragraph{Component power consumption.}
The workflow was executed on cluster nodes equipped with an Intel Xeon E3-1230 V2 processor.
We assume this processor draws 64 W at 50\% utilisation following Equation~\ref{eq:power-dyn}\footnote{ServeTheHome reports 34 W (static) to 94 W (peak) power consumption measured with a power meter for the Intel Xeon E3-1230 V2 processor, \url{https://www.servethehome.com/intel-xeon-e31230-v2-ivy-bridge-xeon-review-4c8t-33ghz/}, accessed 2023-09-30
}.
Following the CCF methodology, we put the power consumption of the DDR3 SDRAM memory to 0.392 W/GB.
As the cluster nodes were only used for this workflow during the experiment, we can attribute the whole 16~GB main memory available per node.
Further following the methodology, we assume that the power consumption for each of the three HDDs per node is around 6.5 W and that power consumed for networking within the data centre can safely be ignored.
For a runtime of 315 minutes on 21 cluster nodes, this adds up to 7.06 kWh for the CPUs, 0.691 kWh for the main memory, and 2.14 kWh for the disks.

\paragraph{Overall power consumption.}
Summing the power consumption of individual components yields a total of 9.90 kWh.
To account for adjacent power consumption in the data centre, such as cooling, we factor in the data centre's PUE. 
Without knowing the specific data centre's PUE, we use an average estimate of 1.6\footnote{\url{https://uptimeinstitute.com/resources/asset/2021-data-center-industry-survey}, accessed 2024-02-26}.
This gives us an estimated power consumption of 15.8 kWh.

\paragraph{Operational emissions.}
The experiment execution was performed in 2021 in the data centre of TU Berlin, located in Germany. 
The average carbon intensity of Germany's energy mix in 2021 was 439 g $CO_{2}e$/kWh.
Therefore, we estimate around 6.95 kg $CO_{2}e$ emissions for running the workflow as described on 21 cluster nodes based on Equation~\ref{eq:emissions}.

The execution was repeated to report a median runtime over three runs.
Hence, we can estimate that around 20.8 kg $CO_{2}e$ were caused to measure the workflow execution three times on 21 nodes.
For comparison, the US EPA's Greenhouse Gas Equivalencies Calculator\footnote{\url{https://www.epa.gov/energy/greenhouse-gas-equivalencies-calculator}, accessed 2023-09-30} translates this to around 53 miles driven in an average gasoline-powered car, a distance similar to driving from San Francisco to San Jose in California.
We note here that the paper also reports workflow executions of seven other cluster scale-outs, so 20.8 kg $CO_{2}e$ are not the total operational emissions of the experiments reported in the paper.

\paragraph{Embodied emissions.}
We can attribute a share of the embodied emissions from manufacturing the cluster nodes to the specific experiment.
The cluster has been in use for around 10 years now.
Hence, with a runtime of 315 minutes, each node was used 0.00599\% of its lifetime for the specific experiment analysed here.
Looking at estimates for the embodied carbon emissions of similar Dell servers compiled by Teads Engineering\footnote{\url{https://docs.google.com/spreadsheets/d/1DqYgQnEDLQVQm5acMAhLgHLD8xXCG9BIrk-_Nv6jF3k/edit#gid=2044051963}}, we can assume an embodied carbon footprint per node of around 1200 kg $CO_{2}e$.
Taking into account the share of the node's lifetime and that the workflow was executed on 21 nodes, this gives us around 1.5 kg $CO_{2}e$ per single execution, or 4.5 kg $CO_{2}e$ to report the median metrics from three executions.
For this estimate, we do not take into account that the nodes were likely reserved for longer, such as also for test runs, and note that we generally do not have information on how long cluster resources were reserved for this or other workloads.

\subsection{An Astronomy Workflow in the Cloud}

\paragraph{Workflow application.}
The Galactic Plane project used the Pegasus workflow system to run the montage workflow and generate around 45 TBs of image mosaics~\cite{rynge_2013_GalacticAWS,Deelman_2016_PegasusCloud}.
They report that running their 16 hierarchical Pegasus workflows took 318,000 core hours to complete using AWS EC2 hi1.4xlarge instances.

\paragraph{Component power consumption.}
The authors did not report the exact runtimes and resource usage and the CCF methodology does not contain a specific data point for hi1.4xlarge instances.
We therefore fall back to general estimates for the energy consumption per core hour.
The CCF methodology puts AWS vCPU core hours to between 0.74 and 3.5 W per core hour and suggests 50\% as reasonable fallback if the exact CPU utilisation is not known, which gives us 2.12 W per core hour or around 674 kWh CPU energy consumption for the the entire workflow.

Looking at the power consumption estimates for AWS EC2 instances compiled by Teads Engineering\footnote{\url{https://engineering.teads.com/sustainability/carbon-footprint-estimator-for-aws-instances/}, accessed 2023-09-30}, 2.12 W per core hour could be slightly too low, but should suffice for a rough estimate.
Their dataset contains the h1.4xlarge instance type, which offers similar processing capabilities to the hi1.4xlarge type used, and consumes about 33\% more power in comparison.

\paragraph{Overall power consumption.}
The authors did not report the allocated or used memory or disk resources during the execution of the workflow application.
We thus take 674 kWh as an estimate of the total energy consumption.
The Teads data puts the PUE of AWS data centres at around 1.2 based on data published by Amazon~\footnote{\url{https://sustainability.aboutamazon.com/products-services/aws-cloud}}.
Therefore, the overall energy consumption should be around 809 kWh -- or more, seeing that we only estimated the energy consumption of the CPUs but not of the allocated memory or disks.

\paragraph{Operational emissions.}
Using the US EPA's Greenhouse Gas Equivalencies Calculator, our estimate translates to around 350 kg $CO_{2}e$ greenhouse gas emissions overall based on a carbon intensity of around 433 g $CO_{2}e$/kWh for the 2019 US energy mix.
This is a similar amount of emissions as three direct flights from Glasgow to Berlin\footnote{\url{https://www.ecopassenger.org} puts a one-way direct flight from Glasgow to Berlin to 114.7 kg $CO_{2}e$ emissions on 2023-09-30}.

\paragraph{Long-term disk storage emissions.}

We know the overall output was 45 TB of Galactic Plane mosaic images.
Assuming these are stored on hi1.4xlarge instances, equipped with local SSD storage, we can use the CCF methodology coefficient for SSDs: 1.2 Wh per TB and hour of use.
Hence, each year of storing these images on SSDs since the paper was published in 2013 can be estimated to correspond to around 473 kWh of energy consumption. With PUE, this grows to around 568 kWh, which translates to 246 kg $CO_{2}e$ emissions per year based on the 2019 US energy mix.

However, in practice, only ``hot'' (frequently accessed) data is kept on disk, and ``cold'' data is moved to tape in data centres. Tape media has a life span of more than 10 years, during which time it can be linked to as little as 5.13 kg $CO_{2}e$ in total emissions (embodied plus operational emissions) per year for storing 45 TB\footnote{Fujifilm estimates the total emissions from storing data on the LTO 8 to be around 0.114 kg $CO_{2}e$ per TB, \url{https://asset.fujifilm.com/www/sg/files/2021-09/d9c014a35ae86bdc41d78abf6e693bb1/Improving_IT_Sustainability_with_Tape_BJC.pdf}}. 

On the other hand, the embodied carbon of 45 TB of SSD capacity, assuming it is required for temporary storage of the images, would amount to around 4.95 t $CO_{2}e$, based on the life-cycle assessment of a Dell server\footnote{The LCA for the Dell R740 server estimates 422.37 kg $CO_{2}e$ for 3.84 TB SSDs, \url{https://www.delltechnologies.com/asset/en-us/products/servers/technical-support/Full_LCA_Dell_R740.pdf}}.

These estimates do not assume data replication. Nevertheless, it is clear that even averaged over a typical 4 to 6 year lifespan, the embodied carbon of the disks will dominate over operational emissions if data is not moved to tape storage.

\subsection{A Bioinformatics Task on Different Infrastructures}
\label{sec:lotaru_example_footprint}

\paragraph{Workflow task.}
We published a trace dataset detailing the execution of popular bioinformatics tasks on a commodity cluster and public cloud resources\footnote{\url{https://github.com/CRC-FONDA/Lotaru-traces}}~\cite{Bader_Lotaru_2022}.
Our trace includes CPU and memory utilisation as well as runtimes of nf-core workflow tasks.
This allows us to compare the energy consumption of a workflow task running on dedicated cluster nodes with that of the task running on public cloud VMs.
For this, we look at the FastQC task.
FastQC is a widely used bioinformatics tool to perform quality control checks on raw sequencing data.
It is used by 74\% workflows of the community-curated Nextflow core-library~\cite{Bader_2024_Lotaru}.
As part of the methylseq workflow, a specific nf-core pipeline, FastQC was run on 36,742,400 sequences in our trace.

\paragraph{Component power consumption.}
On the same cluster at TU Berlin that was used for the Geoscience workflow above, two virtual cores on one cluster node ran FastQC on the input sequences in 155.4 seconds.
The utilisation of the two allocated virtual cores was at 83.75\%.
As the Intel Xeon E3-1230 V2 offers eight virtual cores, leaving six for other tasks, we can attribute 25\% of the processor's static power consumption of 34 W to the task, which works out to around 8.5 W.
Similarly, assuming a dynamic power draw range of 60 W for all eight cores, we can expect the task to draw around 12.56 W for the task's utilisation of the two allocated virtual cores.
Summing this up and factoring in the task's runtime, this gives us an energy consumption of around 0.91 Wh.

GCP's \emph{n2} VM instance type not only offers a similar capacity to the cluster node in terms of cores and memory, but also a similar CPU and memory performance, according to the measurements we obtained using SysBench~\cite{Bader_Lotaru_2022}.
This is further confirmed by the similar runtime, when using two virtual cores on an n2 instance for FastQC: 157.4 seconds for the same input data, so just two seconds slower than on the cluster node.
Meanwhile, the per core utilisation was slightly higher at 94.15\%.
The CCF methodology estimates the power draw of GCP vCores of n2 instances to be between 0.63 and 3.64 W\footnote{\url{https://github.com/cloud-carbon-footprint/ccf-coefficients/}}. 
For two virtual cores utilised as for our FastQC task, the power consumption should, therefore, be around 6.93 W.
Factoring in the task's runtime, this results in an estimated energy consumption of 0.30 Wh.

Seeing that the task memory allocation and the CCF coefficient for memory is the same for both the cluster node and the cloud VM, while the task runtime is also comparable, there is little need to further compare the energy consumption for the allocated memory.
Similarly, the CCF coefficient for disk usage is also based on a general average for data centres.
Therefore, we conclude here that the major difference is in power consumption for processing.

\paragraph{Overall power consumption.}
The significant difference in power consumption for processing –– estimated to be around three times higher for the older cluster node's processor compared to the cloud VM (0.91 Wh versus 0.30 Wh) -- is exacerbated when PUE is taken into account.
For the university data centre we again estimate PUE to be around 1.6, whereas in Google's data centre it should be close to 1.1\footnote{\url{https://www.google.com/about/datacenters/efficiency/}, accessed 2024-03-09}.
Hence, taking PUE into account, processing using the cloud VM can be estimated to be almost 4.5 times more energy efficient for this FastQC task instance: 1.46 Wh (cluster) versus 0.33 Wh (GCP).

\paragraph{Operational emissions.}
Since both task executions were performed in Germany in 2022, we apply the same carbon intensity conversion factor of 473 g $CO_{2}e$/kWh.
Therefore, estimating just the energy consumption of processing, with PUE taken into account, comes out at around 0.69 g $CO_{2}e$ on the cluster and 0.16 g $CO_{2}e$ on GCP.

\paragraph{Embodied emissions.}
Without more details of how long resources were reserved and how resources were shared by multiple tasks, we cannot meaningfully estimate the embodied carbon emissions attributable to the FastQC task.
However, we can conclude this comparison of FastQC on a cluster node and a cloud VM by pointing out that the CCF methodology assumes a lifespan of cloud resources of 4 years, whereas the TU Berlin cluster has been running for around 10 years, so that the attributable embodied carbon is likely substantially lower for the long-running cluster.

\section{Techniques for Individual Tasks}
\label{sec:individual_task_techniques} At the level of the individual
workflow task, reduction of carbon emissions depends on two factors:
the embodied carbon of the hardware on which the task runs, and the
emissions resulting from the energy used for running the task. In
scientific computing, the task runtime energy consumption dominates carbon emissions,
provided that the system is used for several years at full load. This
holds true even in countries like the UK where almost 50\% of electricity
is generated by renewables\footnote{\url{https://www.nationalgrid.com/stories/energy-explained/how-much-uks-energy-renewable}, assessed 2024-07-27}.
We therefore focus on the runtime energy consumption of a given task and its corresponding carbon footprint. Optimising energy efficiency a.k.a \emph{performance per Watt} for a given task means to achieve the lowest possible execution time for a given power budget. With that definition, optimising performance per Watt will lead to the lowest possible energy consumption for a given hardware configuration. This is not only the case on a per-thread basis but also when using hardware parallelism, and also holds for systems with accelerators and for cluster-level parallelism.

\begin{figure}[t]
\includegraphics[width=0.8\textwidth]{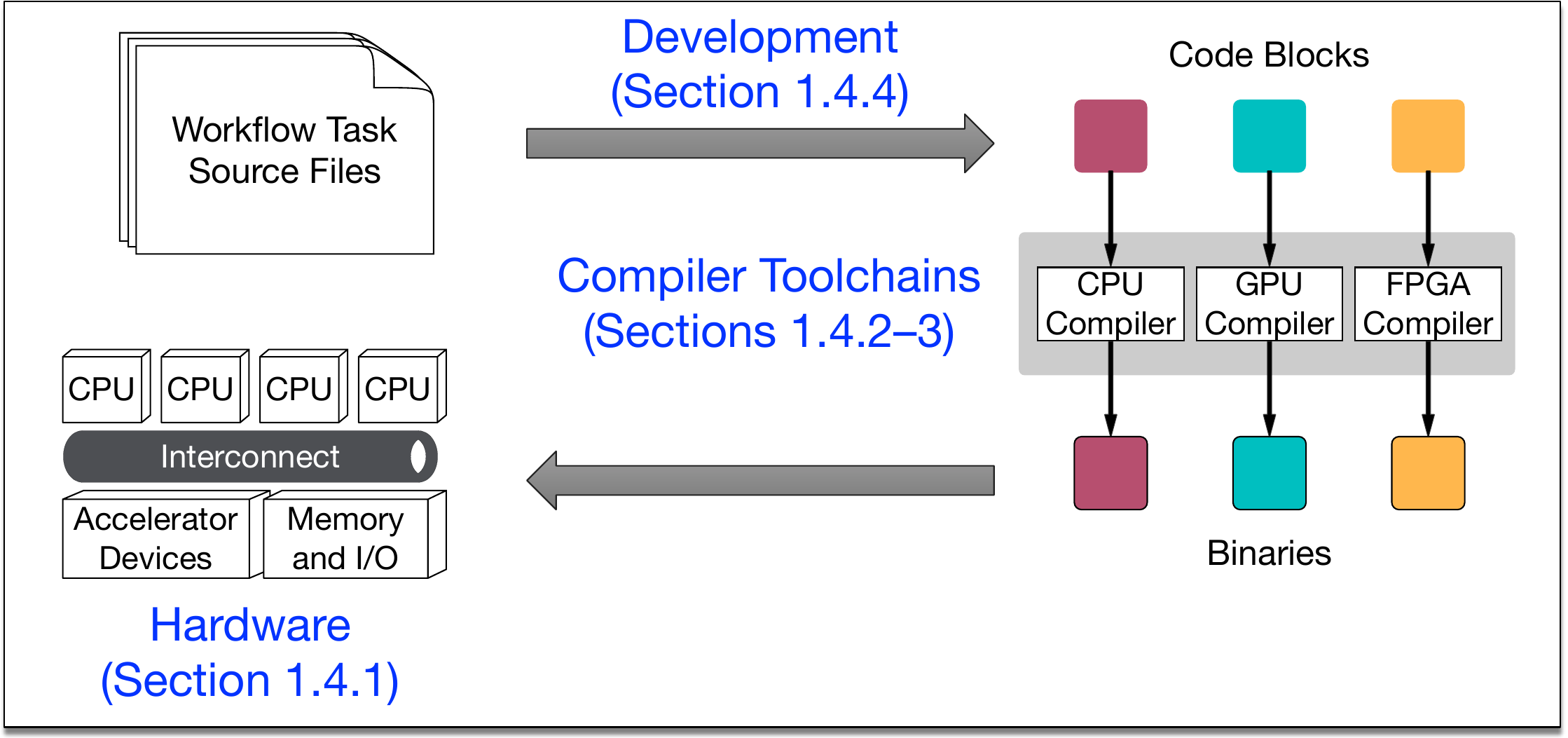}
\centering \label{fig:task-optimisation-intro} \caption{An individual workflow task, divided into blocks of code, is partitioned to and optimised for execution on different types of devices on a single compute node.}
\end{figure}

However, embodied carbon also represents a significant fraction of
CO$_{2}$ emissions of a compute system and must be considered as well. For most scientific computing use cases, the key consideration
is the amount of disk space required for a task, as disks are typically responsible
for the majority of the embodied carbon in a server. To some extent
this is under the control of the task designer (i.e.\ the scientific
domain expert who creates the code and runs the experiments or simulations).
However, such users of a cluster may not have a say in its specification
and procurement. The amount of memory used by a task also contributes,
especially in cases where tasks need to be distributed over many nodes
in a cluster because of their memory needs. In this case, using more
memory leads to additional energy consumption. Therefore, saving memory
on distributed compute tasks is crucial. Since for many tasks the
bottleneck is memory bandwidth, reducing memory usage can directly
improve execution speed, which in turn translates to better performance per Watt. %
Figure~\ref{fig:task-optimisation-intro} provides an overview of
the optimisation techniques related to single tasks discussed in this
chapter. 

\emph{Energy-Efficient Hardware} (Section~\ref{sec:Energy-Efficient-Hardware}) refers to the choice between different heterogeneous computing devices, with implications for energy-efficiency (performance per Watt) for a given task.

\emph{Energy-Efficient Compilers: An Overview} (Section~\ref{sec:Energy-Efficient-compilers-overview}) describes different routes to task programming and compiling, device-specific programming, domain-specific programming, pragma-based approaches, and compilation from legacy code.

\emph{Energy-Efficient Compilers: Techniques} (Section~\ref{sec:Energy-Efficient-compilers-techs}) summarises specific transformation techniques used by compilers to optimise programs, with a focus on different types of cost-performance models for exploring the domain of transformations.

\emph{Energy-Efficient Development and Deployment} (Section~\ref{sec:Energy-Efficient-dev-and-dep}) concludes the section with a brief discussion of the impact of and suggested approaches to energy-efficient development and deployment of computing tasks.

\subsection{Energy-Efficient Hardware}
\label{sec:Energy-Efficient-Hardware}

For every task there is an optimal platform in terms of energy efficiency.
For some tasks this will be the CPU, for others accelerators such
as GPU, FPGA, TPU, and NPU.  Liu and Luk, for instance, investigated CPUs, GPUs, and FPGAs in the context of energy-efficient scientific computing, and highlighted the sensitivity of system energy efficiency to the choice of individual computing components ~\cite{10.1007/978-3-642-28365-9_6}. Therefore, energy-efficient systems for heterogeneous workloads must also be heterogeneous themselves.

At the same time, within each of these categories there is a wide
range of architectures and hardware system designers need to make
decisions on the hardware mix based on various trade-offs. 
The deployed
system also needs to be managed dynamically to find the optimal configuration
of the system itself for a given task, e.g. CPU voltage and/or frequency. 

Accelerator-based heterogeneous systems do require careful tuning for specific patterns of computation, and are better suited for recurring and predictable workloads.
There is however overwhelming evidence for the potential benefit of accelerators
for specific application domains, in particular scientific computing
and machine learning. As a result, the use of accelerators has been
increasing: 169 of the 500 fastest computers in the top500 list use
accelerators\footnote{https://www.top500.org/lists/top500/2022/06/highs/};
more significantly, each of the top 10 in the list of \emph{green}
supercomputers uses an accelerator\footnote{https://www.top500.org/lists/green500/2023/11/}.

\paragraph{Accelerators.} GPUs can accelerate Earth science models that incorporate
Deep Learning~\cite{wang2023adopting}; our work~\cite{6641457} has
shown the potential of GPUs for acceleration of Numerical Weather
Prediction and similar fluid dynamics simulation workloads. FPGAs
have been receiving increasing attention, especially since being made
available in the Amazon cloud. In their review paper, Bobda et al.
identify deep learning and scientific computing as promising workloads
for FPGAs in data centres~\cite{bobda2022future}. Our own work has
also demonstrated the potential of FPGAs for scientific computing
\cite{8994801} and information retrieval tasks~\cite{5272246}, the
latter is further evidenced by the now extensive use of FPGAs for Microsoft's
Bing search engine. TPUs were specifically designed as accelerators
for Deep Learning workloads, however a recent comparison with GPUs
and CPUs~\cite{9691951} shows that the advantage depends a lot on
the specific Deep Learning workload.

\paragraph{CPUs.} Low-power RISC CPU cores offer the potential for improved energy efficiency
for HPC workloads. Arm nodes have been deployed in prototype compute
clusters like Mont Blanc~\cite{rajovic2016mont}, and also in production
high-performance systems like Isambard~\cite{mcintosh2019performance}.
Although these RISC CPUs generally have lower single core performance,
their relative simplicity means core density can be higher which might
result in improved parallel throughput.

Other RISC-based hardware such as the open source RISC-V initiative
might be suitable for HPC~\cite{berger2023evaluation} although it
is apparent that fabrication, integration, and toolchain support are
relatively immature in contrast to Arm.

\subsection{Energy-Efficient Compilers: An Overview}
\label{sec:Energy-Efficient-compilers-overview}

The purpose of a compiler is to turn source code into an executable
format. The additional purpose of an energy-efficient compiler is
to ensure that the energy for executing the program is minimised.
Even ordinary compilers have a huge effect on the energy efficiency
of programs because in general, faster single-threaded execution means lower energy
consumption. This is even the case for interpreted languages like
Python because for compute-intensive operations, such languages typically
call on compiled libraries of code written in C, C++, or Fortran.

Current compilers are not sophisticated enough to optimise or even
compile a given code to arbitrary heterogeneous systems. This requires
a combination of automated program transformation (typically for transformation
of parallelism, but also e.g. to trade memory for computation or vice
versa) and cost-performance models, as well as formalised representations
and rewrite rules. There is a large body of research on compilation
to energy-efficient accelerators such as GPUs and FPGAs. In this research,
there are several categories depending on the approach. Essentially,
they differ in how much effort they expect from the programmer.

\paragraph{Custom accelerator programming languages.} The most popular approach has been to define a custom language to
program the accelerator. For GPUs, typical examples are OpenGL and
CUDA. For FPGAs, there are too many to enumerate, but noteworthy examples
are Bluespec\footnote{\url{https://github.com/B-Lang-org/bsc}} and
Vivado HLS\footnote{\url{https://www.xilinx.com/products/design-tools/vivado/high-level-design.html}}.
The former is a superset of SystemVerilog, which is a superset of
Verilog, one of the most popular Hardware Design Languages (HDL).
This approach is based on raising the abstraction level of HDLs. The
other languages, including the GPU languages, are C-based, and for
FPGAs they are colloquially known as ``C-to-gates'' languages. Because
C is intrinsically a language for a sequential von Neuman machine,
extracting parallelism is difficult. Therefore, such languages require
the programmer to be explicit about parallelism. For GPUs, this typically
means data parallelism. For FPGAs, the matter is more complicated as
they will usually perform better with pipeline and task-level parallelism,
and those are even harder to express in a C-based language. As a consequence,
languages such as OpenCL and SYCL provide additional constructs to
express dataflow pipelines of computation.

A different approach to custom language design is to use a language,
typically a functional language, that has parallelism by default and
is designed specifically for compilation to accelerators. For example,
Lift~\cite{steuwer2017lift} provides rewrite rules to target various
platforms. This means the programmer does not have to concern themselves
so much with the details of parallelism. However, they have to learn
a specific language which is very different from ordinary C-style
languages.

\paragraph{Pragma-based approaches.}

For a long time, OpenMP has provided a popular approach to parallelisation
of CPU code by adding special annotations, known as \emph{pragmas},
to the source code. The idea is that on a sequential system, such
pragmas are ignored, and on a parallel system, they are used by the
compiler to create parallel code. In practice, there is a need for
an additional API to handle resource contention, so usually OpenMP
code is a mixture of pragmas and dedicated API calls. More recently,
there have been efforts to create a similar approach to accelerator
programming, known as OpenACC. At the same time, the OpenMP specification
has been extended to include accelerator offloading. The perceived
advantage of pragma-based approaches is simplicity, because the original
sequential code is only annotated. In practice, to achieve good performance,
either the code needs rewriting or the pragma-based annotations become
quite long and complicated, so that the advantage in terms of effort
over CUDA, OpenCL, or SYCL is small.

\paragraph{Domain-specific accelerator programming languages.}

Another recent trend is to use an embedded domain-specific language
to program accelerators. These have the advantage that, provided they
know the host language, the programmer does not need to learn a new
language, only a new API within that language. The key examples of
this approach are Chisel~\cite{bachrach2012chisel} for FPGAs and
Obsidian~\cite{svensson2008obsidian} for GPUs. There are also many
domain-specific languanges, and some of these target multiple accelerator
platforms, for example HeteroHalide~\cite{li2020heterohalide} for
image processing, TensorFlow~\cite{pang2020deep} for machine learning,
etc. These are cross-platform as teams of experts develop, tune, and
package libraries for different platforms.

This approach is mostly made possible by research on platform-independent
accelerator languages such as OpenCL and SYCL. All of the above approaches
have the disadvantage that they require to write new code, because
they cannot be used to automatically compile legacy code to accelerators.
Furthermore, they are in general not performance-portable, so the
code needs to be rewritten for optimal performance on a new target
architecture.

\paragraph{Compilation from legacy code.}

There is active research on compilation
from legacy code. Our work~\cite{vanderbauwhede2018domain} shows
that legacy code can be made type-safe and auto-parallelised to GPUs
and FPGAs; our approach of type-driven transformations also facilitates
many optimisations, e.g. trade memory for computation~\cite{10035181}.

The work by Vandebon et al.~\cite{vandebon2022meta} follows a meta-programming
approach to optimise for different heterogeneous targets. This involves
writing optimisation strategies in a Python-based embedded domain-sepcific
language. The compiler uses those strategies to optimise the original
program. The work by O'Boyle et al.~\cite{10.1145/3578360.3580262}
uses machine learning to detect specific algorithms (FFT, matmult,
convolution) and replace them with calls to fast accelerator libraries.

\subsection{Energy-Efficient Compilers: Techniques }
\label{sec:Energy-Efficient-compilers-techs}

Simplifying slightly, we can say that a resource-unaware compiler
translates from source code to binary format. A resource-aware compiler
transforms the program to meet the resource constraints. If the resource
to optimise for is energy, then we have an energy-efficient compiler.

In practice, all compilers perform some degree of program transformation,
such as register optimisation, spilling, inlining, loop unrolling,
and auto-parallelisation. In current compilers, such transformations
use ad-hoc cost-performance models, typically at the level of the
cost of executing a given instruction, with the aim of optimising
for performance, binary compatibility, or executable size. However,
to obtain truly energy-efficient compilers, program transformations
need to be much more comprehensive, and crucially, they need a much
deeper understanding of the target platform than current compilers
have. We will briefly discuss various techniques for program transformation
as well as the cost-performance models that are needed to be able
to determine the optimal transformation.

\subsubsection{Program Transformation Techniques}

Higher-level program transformation techniques rely on a technique
known as static analysis~\cite{thomson2021static}. This means analysing
an abstract representation of the parsed source code to identify portions
of the program that should be rewritten. The challenge for the actual
rewrite of the program is to ensure not only better energy-efficiency,
but also correctness. It is easy to transform a program, but for automated
program transformation we need a guarantee that a sequence of transformations
results in correct code. 

The most well-known example of a framework for program transformation
is probably the Polyhedral Model~\cite{trifunovic2009polyhedral},
which allows auto-parallelisation of complex loop nests. Another well-known
approach is the use of rewriting systems, either term-based~\cite{schlaak2022memory},
pattern based as in MLIR~\cite{lattner2021mlir}, or type-driven as
in our work~\cite{vanderbauwhede2019type}.

\subsubsection{Cost-Performance Models for Performance-Portable Compilation}

Energy-efficient compilers need to have a cost-performance model that
enables exploration of possible transformations (design-space exploration,
DSE). The `cost' predicted by such models can relate to different
metrics, e.g., execution time for a given power budget (performance per Watt), memory use, code size,
programming effort, or energy consumption. 

Cost models can operate at different levels in a workflow system.
In this section, we are focused on cost models for the compilation of tasks for
single-node targets, potentially with accelerators (Section~\ref{sub:modelling}
discusses entire workflow modelling). Such cost models aim to identify
``hot code" -- the relatively few blocks of code that take dominate the execution time -- and optimise them for target hardware, which
can be expected to be parallel, and increasingly heterogeneous. The
complexity of modern software and hardware makes the design of cost
models a non-trivial task, and because a brute-force approach is often
unrealistic in a large design-space, heuristics and ML models are
often employed, such as discussed in~\cite{siraichi2016design}.

\paragraph{Models for data transfer.}

A heterogeneous system consists of several compute nodes that can
communicate with one another. In current architectures, the typical
model is a host CPU with one or more accelerators connected via PCIexpress.
The key feature is that typically, the accelerators have their own
memory, and can only access the host memory indirectly. Therefore,
in general a lot of data needs to be copied back and forth between
the host and the accelerator. It is relatively easy to devise a first-order
cost-performance model for the cost of the data transfer to decide
if offloading to the accelerator is worthwhile, see e.g.~\cite{6641457}.
More complex cost models for data transfer take into account access
patterns. In practice, it is not feasible to do this entirely analytically:
the performance of various access patterns instead needs to be obtained
from memory test suites such as MP-STREAM~\cite{nabi2018mp}.

\paragraph{Models for memory access.}

Cost models also require a \emph{memory} model. A key challenge is
the probabilistic nature of caches in a typical multi-level memory
architecture of CPUs and GPUs. Our work on modelling memory behaviour
in heterogeneous devices proposes an empirical approach~\cite{nabi2018mp}.
Other sophisticated approaches have also been proposed, some of them
are discussed in~\cite{4228126} and~\cite{zhong2009program}.

\paragraph{Models for CPU-based computations.}

The compiler goes through a number of stages to translate source code
into executable machine code. The optimisation stage is typically
where cost models are employed to inform the DSE so that the right
set of transformations is applied, for instance to maximize energy efficiency.
As the design space keeps evolving and expanding, with parallelism
at multiple levels, determining the optimal sequence of transformations
becomes a non-trivial task, even when supported by cost models.

There is considerable work in the literature proposing more sophisticated
cost models beyond the ad-hoc models built into existing compilers,
e.g.~\cite{10.1007/978-3-540-89740-8_16,7429329,10.1145/3281287.3281290}.
Parallel programming libraries such as OpenMP can also have associated
cost models~\cite{4228126}.

\paragraph{Models for accelerator-based computations.}

There is prior work on the use of cost models for GPUs, e.g.~\cite{mishra2022compoff}.
FPGAs are also increasingly used as accelerators. As their compilation
times are very long, cost models that can quickly predict cost of
design variants are especially useful. Predictions of off-the-shelf
FPGA compilers are too slow to carry out meaningful DSE; this limitation
can be addressed by building light-weight cost models~\cite{nabi2019fpga}. 

\subsection{Energy-Efficient Development and Deployment}
\label{sec:Energy-Efficient-dev-and-dep}

According to our study~\cite{wim_vanderbauwhede_2023_7709483} on
the impact of the development and deployment of scientific software
on energy consumption, for mainstream scientific computing tasks,
energy is often wasted because the development and deployment processes
are sub-optimal, e.g. lack of auditing and sign-off process, continuous
integration with unit testing, integration testing, and acceptance
testing. There is also often insufficient knowledge of optimal use
of programming languages, heterogeneous computing devices, and compilation options.

\paragraph{Impact of development and deployment on energy consumption.}

We have presented detailed recommendations on minimising full-lifecycle
emissions of institutional facilities and on optimising code and reducing
coding mistakes~\cite{wim_vanderbauwhede_2023_7709483}. In summary,
these involve 
\begin{itemize}
\item institutional/facility-level policies, such as processes for testing
and deployment or policies to address CPU/memory hogging and manage
disk space; 
\item expert support in green software engineering and energy-efficient
software design; 
\item and provide adequate training for researchers as well as experts. 
\end{itemize}

\paragraph{Approaches to improve software energy efficiency.}

In~\cite{wim_vanderbauwhede_2023_7709401} we presented an overview
of the state of the art on several approaches to improve software
energy efficiency: 
\begin{itemize}
\item code optimisation to speed up single-threaded execution and to reduce memory access; 
\item parallelisation across multiple hardware cores to speed up execution; 
\item tuning compiler optimisations; 
\item and code rewriting to target accelerators. 
\end{itemize}
A key observation is that there is no single win in terms of improving
software energy efficiency: all of the above approaches need to be
considered, and require expert knowledge. 

\section{Techniques for Entire Workflows}
\label{sec:entire_workflow_techniques}

\begin{figure}[t]
\includegraphics[width=0.8\textwidth]{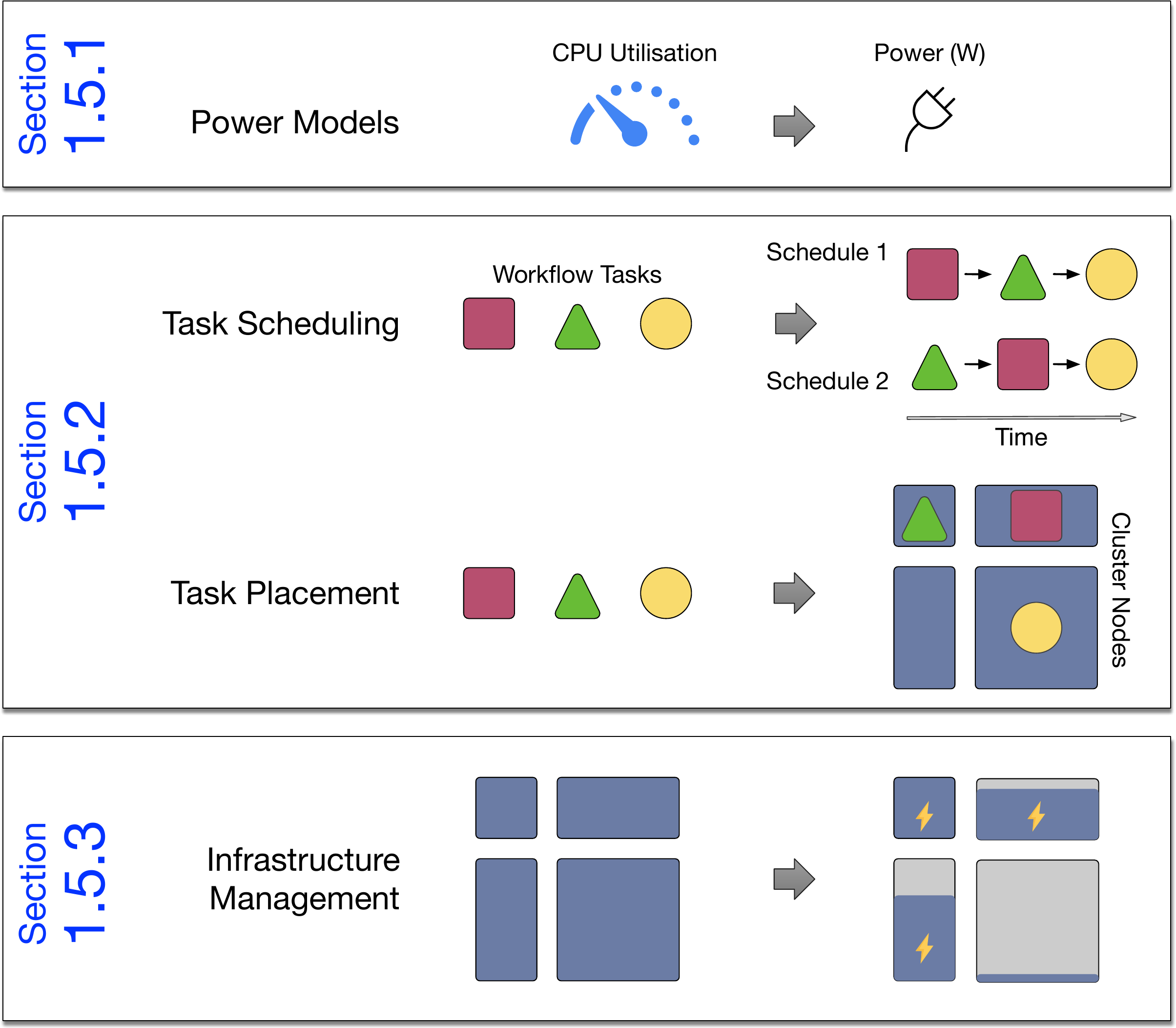}
\centering
\label{fig:scheduling-intro}
\caption{Task power modelling and estimation, task scheduling and placement, and infrastructure management for energy-efficient workflow execution as presented in this section.}
\end{figure}

Having considered the energy efficiency of individual tasks in Section~\ref{sec:individual_task_techniques}, we now focus on techniques to reduce the environmental footprint of entire workflow applications.
The techniques we describe fall into three broad categories: \emph{task estimation}, \emph{task scheduling and placement}, and \emph{infrastructure management}. Figure~\ref{fig:scheduling-intro} provides an overview of these categories of cluster resource management techniques.

\emph{Task modelling and estimation} (Section~\ref{sub:modelling}) is the process of characterizing the performance, resource utilisation, and energy consumption of executing a task on a given compute node. It may consider metrics, such as CPU cycles, RAM usage, I/O utilisation, proximity to other tasks, and energy consumption in producing a `score'. This score can then be used to compare options for task \textit{placement} and \textit{scheduling}.   

\emph{Task scheduling} (Section~\ref{sub:sched-place}) refers to the process of determining the order in which a workflow's tasks are executed. This considers precedence constraints between the tasks, as well as possible parallel execution of tasks within a workflow. \emph{Task placement} (Section~\ref{sub:sched-place}) refers to the process of determining which compute node a task should be placed on to execute. Task scheduling is often performed by Scientific Workflow Management Systems (SWMSs), while task placement is usually performed by cluster resource managers~\cite{lehmann2023cwi}. However, many energy-aware workflow scheduling methods determine both the order tasks get executed in, as well as their placement onto cluster nodes. 

\emph{Infrastructure management} (Section~\ref{sub:infra-man}) refers to the way in which the compute cluster infrastructure is configured. Examples can include the amount of RAM, CPU frequency, or network bandwidth made available to a task. In virtualised compute infrastructure, such as cloud environments, infrastructure management can include the selection of the types, sizes, and amounts of virtual machines to be provisioned for a workflow application. 

\subsection{Task Energy Efficiency Modelling and Estimation}
\label{sub:modelling}

To manage cluster resources for a workflow in an energy-efficient manner, it is important to understand the effect that resource allocations have on a task's runtime, resource utilisation, and energy consumption. 
To optimise allocations based on this effect, energy-aware scheduling and placement methods model and estimate it.

\paragraph{Task energy consumption.} The energy consumption of a workflow task can be estimated based on the cluster resources allocated and used to execute the task, including a cluster node's CPU, RAM, and storage.
It is important to note that these components consume different amounts of energy depending on their level of utilisation, as explained in Section~\ref{sec:computing_footprint}. 
For example, a CPU-bound task and an I/O-bound task that execute for the same amount of time on the same node will tend to consume different amounts of energy.

At any given moment, the power consumption of a workflow task executing on a cluster node can be summarised in terms of Equation~\ref{eq:power-total} in Section~\ref{sec:footprint_estimation_methodology}: 
When executing workflow tasks on a cluster node, $P_{static}$ refers to the minimum power required to run the node, and $P_{dynamic}$ is the power required to run one or more tasks.
If a cluster consists of heterogeneous nodes, the power consumption of a task also depends on the specific node upon which it executes, meaning that $ P_{static}$ and $ P_{dynamic}$ will not be constant across heterogeneous cluster nodes.
Similarly, also a task's runtime and resource utilisation will depend on the specific node used to run the task in a heterogeneous cluster.
Another important factor is whether other tasks share a node's resources, so that multiple tasks split the static energy consumption required to have the node be operational.

\paragraph{Task power models.}
Different approaches are used to model the energy consumption of tasks executing on a node. The simplest is to assume a linear relationship between a node's resource utilisation and energy consumed.
Creating such linear power models can be accomplished with only two data points, namely the static energy consumption of idle resources and the dynamic energy consumption of a given level of resource utilisation. These models can be constructed using data points reported by manufacturers, measured using software or hardware energy meters, or obtained from literature or online sources. This makes them easy to implement and interpret.
In addition to linear power models, non-linear models~\cite{JIN2020114806} and neural networks~\cite{9302777} have also been used to translate resource utilisation into energy consumption.

\paragraph{Models for workflow scheduling.}
Power models used for energy-aware workflow scheduling differ significantly in what resources they include. 
Many approaches~\cite{yassa2013multi,Pietri_2013_SPSS-EB,Durillo2015moheft,Ghose2017energy,Li_2018_CEAS} focus on mapping CPU utilisation to energy consumption. 
These tend to use models that linearly interpolate between the energy consumption levels at the idle state and peak CPU load, and this has been found to work well for CPU-intensive workloads.
However, for workloads that are I/O-intensive, approaches that only take CPU utilisation into account may result in inaccurate assessments~\cite{dasilva2020char}.
Hence, models that also account for the energy consumption of storage~\cite{Ferreira2019simulating,dasilva2020char,Coleman_2021_IOWorkflowPower} and networks~\cite{Bousselmi2016wpem} have been proposed.

Some studies assume that task energy consumption across the heterogeneous nodes of a cluster can be measured to inform power estimation models.
One way to approach this is to employ software energy meters (such as the RAPL interface on Intel/AMD processors\footnote{\url{https://www.intel.com/content/www/us/en/developer/articles/technical/software-security-guidance/advisory-guidance/running-average-power-limit-energy-reporting.html}, accessed 2024-05-07} and NVML\footnote{\url{https://developer.nvidia.com/management-library-nvml}, accessed 2024-05-07} on NVIDIA GPU devices) while recurring tasks are executed on a variety of cluster nodes.
If many data points have been collected this way, power models can be more complex than models that map resource utilisation to energy consumption.
For example, Durillo et al.~\cite{DURILLO2014221} train neural networks to estimate task energy consumption based on the input size, task parameters, and information on co-located tasks.

\subsection{Energy-Efficient Task Scheduling and Placement}
\label{sub:sched-place}
Energy-aware workflow scheduling has been studied extensively~\cite{Saurav2021survey}.
The central idea is to assign workflow tasks to nodes in a way that minimises energy consumption.
This is possible because nodes in a cluster are often heterogeneous and provide different levels of energy efficiency in addition to the heterogeneity of workflow tasks themselves.
For instance, some tasks might be CPU-intensive, while others predominantly rely on I/O.
This creates opportunities to schedule the various tasks of a workflow application to the nodes of a heterogeneous cluster in such a way that minimises overall energy consumption.
Moreover, there are often additional objectives and constraints for task scheduling and placement, e.g. to ensure low costs or that a deadline is met.

\paragraph{Energy-aware HEFT variants.} Widely recognised works on energy-aware workflow scheduling include GreenHEFT~\cite{DURILLO2014221} and MOHEFT~\cite{Durillo2015moheft}.
Both are greedy heuristic scheduling algorithms inspired by the seminal HEFT algorithm~\cite{Topcuoglu_2002_HEFT} for scheduling workflow tasks on heterogeneous cluster nodes. %
Similar to HEFT, GreenHEFT and MOHEFT have two phases: ranking and mapping. 
In the ranking phase, tasks are ordered according to their distance from the end of the workflow, prioritising the longest chains of consecutive tasks, exactly as in HEFT.
Subsequently, in the mapping phase, where HEFT always maps the next task to be scheduled to the node offering the lowest runtime, GreenHEFT instead maps to the node that offers the lowest energy consumption for the task.
MOHEFT works in a similar way, but it builds multiple solutions simultaneously in its mapping phase, aiming to identify the Pareto front of the best possible trade-offs between runtime and energy consumption.
For the mapping phase, the algorithms require knowledge of the energy consumption and runtime of the tasks on available nodes.
This stems from previous executions of the tasks used to train estimators per task for both runtime and energy consumption.

\paragraph{Other schedulers.} Many other energy-aware workflow schedulers make use of the structure of workflow applications and information of the energy consumption of tasks.
The structure typically takes the form of a directed acyclic dependence graph between tasks, where energy consumption is either measured or estimated.
Commonly, this information is then provided to a heuristic algorithm that schedules the tasks according to precedence and other constraints while minimising energy consumption.
PASTA~\cite{sharifi2013pasta} selects a subset of available resources that can balance between lower overall power consumption and a shorter makespan. 
EnReal~\cite{xu2015enreal} makes decisions based on estimations of the energy cost of each task. 
The work of~\cite{khaleel2016energy} operates similarly, but consolidates at the VM rather than at the task level. 
EONS~\cite{Chen2016eons} selects hosts with the highest ratio of the maximum frequency to power at said frequency. 
\cite{Ghose2017energy} calculate the amount of `slack time' between tasks and make use of it using three alternative heuristics in order to increase VM utilisation. 
PESVMC~\cite{mohanapriya2018energy} finds for each task the VM that has the most resources available and can execute the task with the least amount of energy consumption. For this, it uses estimations of the task performance. 
EEWS~\cite{singh2020makespan} is a metaheuristic algorithm inspired by the Hybrid Chemical Reaction Optimisation algorithm. It iteratively applies chemical reactions to a population of solutions with the goal of finding a solution that optimises the objective function, which in this case is to minimise energy consumption while meeting deadline constraints.
pSciMapper~\cite{zhu2010power} is a power-aware consolidator that aggregates resource usage characteristics of workflow tasks (namely CPU, memory, disk, network) and applies hierarchical clustering to identify interference as an indicator of resource contention.
It then maps clustered task sets to servers to save power and costs by avoiding interference.

\subsection{Energy-Efficient Infrastructure Management}
\label{sub:infra-man}

Complementing the scheduling and placement of workflow tasks, it is often also possible to configure the cluster infrastructure so that workflows execute with a high degree of energy efficiency. This is often done to balance economic, performance, and energy-efficiency objectives. The infrastructure configuration can happen at the physical \textit{hardware} or \textit{virtualised} software infrastructure levels. 

\paragraph{Hardware infrastructure management.} Dynamic voltage and frequency scaling (DVFS)~\cite{weiser1996scheduling} can be used to change the voltage supplied to a CPU at runtime to effect energy consumption and frequency. As a rule of thumb, higher voltage and frequencies result in shorter execution times for CPU-bound workloads, but can be less energy-efficient. In contrast, lower frequencies can be more energy-efficient, especially for I/O-bound workloads. Despite lessening return of DVFS compared to early compute hardware~\cite{10.5555/1924920.1924921}, it remains a popular approach especially for large compute clusters where savings scale with the number of machines.
DVFS can also be employed for RAM, where refresh rates can be relaxed to reduce energy consumption while still maintaining data integrity~\cite{10.1145/2485922.2485928}. Of course, static, pre-defined specification of the compute infrastructure can be used to provide clear bounds on efficiency and performance goals. However, this precludes opportunistic savings at runtime.

\paragraph{Virtualised infrastructure management.} Virtualised software infrastructure provides more flexibility and programmability compared to hardware infrastructure. Here, the configuration and number of virtual machines (or containers) deployed across a cluster can be managed specifically for current tasks. For example, tasks with parallel workloads may benefit from multiple virtual CPUs (vCPUs) per VM, while others may be more suited to fewer but faster vCPUs. Equally, I/O-bound tasks may only need a single, slow vCPUs, but large amounts of local storage or network bandwidth. Furthermore, unused cluster resources can be powered down or released for other workloads to make use of.
Consolidating resource usage onto a subset of resources saves static power and spreads the remaining static power across more tasks, resulting in more energy-proportional computing.

\paragraph{Methods for workflows.} Similarly to Section~\ref{sub:sched-place}, several approaches have been proposed to use infrastructure management techniques to achieve energy efficiency in workflow applications. Using a variant of the crow swarm optimisation algorithm, Reddy and Reddy~\cite{Reddy2023} demonstrate that the selection and deployment of pre-defined VM configurations can be achieved to reduce resource consumption, energy consumption, and execution time based on analysis of workflow characteristics. Medara et. al~\cite{MEDARA2021102323} describe how task scheduling is extended to include the consolidation of VM spread across different nodes to a single node based on the principles of wave propagation, enabling unused nodes to be powered down. CEAS~\cite{Li_2018_CEAS} aims to minimise both cost and energy consumption by selecting from multiple VM instance types, chaining specific tasks in workflows before deployment, and reusing idle VMs.
RMREC~\cite{Zhang2020RMREC} goes further and performs dynamic reallocation of workflow tasks to VMs based on energy consumption, task execution profile, and network bandwidth requirements. It also leverages the migration of VMs between hardware nodes. The energy cost of such migration is included in the cost modelling and decision process. Beyond cost, time, and energy performance, Stavrinides and Karatza~\cite{Stavrinides2019qos} include application-layer quality of service as part of the decision logic in configuring the DVFS of the cluster system for workflows. Here, application accuracy is traded for energy savings.

\section{Discussion}
\label{sec:discussion}

In this section, we look beyond energy-efficient task implementations and workflow scheduling by discussing the impact of compute environments, energy sources, as well as economical and rebound effects.

\subsection{The Impact of Compute Environments}

Looking past the energy efficiency of task codes and cluster nodes, the energy efficiency of compute environments can vary considerably also in terms of adjacent energy needed for cooling, lighting, and other support functions.
As explained in Section~\ref{sec:computing_footprint}, a key indicator of this is PUE, which measures how much energy is needed on top of what is used to directly power computing equipment.
Even though PUE as a simple metric is not without shortcomings, as discussed in the background section, it can help scientists choose a compute infrastructure for their workflows that is operated more efficiently overall.

Resources provided by major cloud providers usually have a significantly better PUE than resources housed in other data centres~\cite{Barroso_2019_DatacenterComputer}, as also exemplified in Section~\ref{sec:lotaru_example_footprint}.
This is due to more energy-efficient operations at scale, with major cloud providers investing significantly in reducing operational costs, where energy costs are typically a major factor.
In addition, it is safe to assume that resource utilisation tends to be higher in clouds. 
One reason for this is that resource pooling typically improves utilisation, and large public cloud infrastructures tend to be shared by more users than smaller private clouds or clusters.
Another reason is that the continuous cost of public cloud resources directly incentivises good use of resources, whereas there is often no perceived cost to allocated but idle private resources.

On the other hand, because clouds promise virtually unlimited resources and guarantee high levels of service, major cloud data centres need to be conservatively sized to cope well even with unusual peaks in demand, and are therefore likely to be significantly underutilised as well~\cite{Barroso_2019_DatacenterComputer}.
The extent of surplus resources in major clouds is indicated by their spot markets, where all major cloud providers offer spare resources at a fraction of on-demand prices, and for many VM instance types, surplus resources are available virtually constantly~\cite{Lee_2022_SpotLake}.
This built-in overcapacity comes with a considerable embodied footprint.
Furthermore, as described in Section~\ref{sec:lotaru_example_footprint}, major cloud hardware is assumed to have a lifetime of around four years, whereas, university clusters can be used for a much longer time, so the amount of embodied carbon attributable to running any single workflow application may be lower.

\textbf{Take away:} Data centre energy efficiency and embodied carbon emissions can vary significantly and must, therefore, be considered alongside code and scheduling efficiency when deciding which compute infrastructure to run scientific workflows on.

\subsection{The Impact of Energy Sources}

The emissions caused by energy consumption depend on the sources of energy.
As explained in Section~\ref{sec:computing_footprint}, carbon intensity as a metric captures this by measuring the emissions per unit of energy consumed from a mix of energy sources.
It tends to vary from one regional grid to the next, as energy systems are powered by different sources, face different demands, and are influenced by local weather conditions~\cite{Wiesner_2021_LetsWaitAwhile,CarbonCast}. 
For the same reasons, the carbon intensity of grids often fluctuates also over time as more or less renewable energy is generated from variable sources, such as wind and solar, against a varying overall demand.
In the UK, for example, wind power generates the majority of renewable energy\footnote{\url{https://www.nationalgrid.com/stories/energy-explained/how-much-uks-energy-renewable}, accessed 2024-05-15} and coincides with low grid demand on windy nights.

Some workflows are executed in public clouds, allowing scientists to choose a particular cloud provider and region not only on the basis of financial costs and PUE, but also based on the carbon intensity of a cloud region's energy mix.
In addition, scientific workflows can be delay tolerant, meaning that results are not necessarily needed as soon as possible, but within a few hours, days, or weeks.
This allows to align workflow energy consumption with the availability of low-carbon energy even within a single region, essentially waiting for low-carbon energy to become available.
This approach, which leverages the variability of low-carbon energy for flexible workloads, has become known as \textit{carbon-aware computing}~\cite{Wiesner_2021_LetsWaitAwhile,Radovanovic_2023_GoogleCarbonAware,Lin_2023_GreenCapacities}.
It has also been investigated for workflow applications, such as BLAST~\cite{Souza_2023_Ecovisor} and nf-core bioinformatics pipelines~\cite{Bader_2024_Lotaru,westCCGrid2025}.

There is an ongoing discussion about the best signal to use for carbon-aware spatio-temporal load shifting.
A key question is whether \emph{average} or \emph{marginal} carbon intensity should be optimised. 
The average carbon intensity reflects the emissions of individual energy sources according to their relative share of the overall energy mix.
Marginal carbon intensity, meanwhile, focuses only on the source of the energy that is added or removed as a consequence of load shifting.
Most of the work so far uses average carbon intensity as it is easier to predict and, therefore, forecasts are more readily available and reliable.
There is, however, also work that uses marginal carbon intensity as a signal~\cite{Huang_2023_CarbonVMs,Maji_2023_MultiCloudCarbon}, and some methods focus specifically on using otherwise curtailed renewable energy~\cite{ZHENG20202208,wiesner2022cucumber,wiesner2024fedzero}.

It is recognised that naive carbon-aware computing methods could lead to new or amplified load peaks, which could drive further investment in additional cloud capacity and, thereby, increase embodied carbon emissions~\cite{10.1145/3575693.3575754}.
Furthermore, naive load shifting at scale also poses a risk to grid stability\footnote{\url{https://hackernoon.com/carbon-aware-computing-next-green-breakthrough-or-new-greenwashing}, accessed 2024-05-10}.
It is therefore crucial that low carbon intensity is not the only optimisation goal.

As grids are decarbonised over the coming decades, the embodied emissions of computing resources will become more dominant than the operational emissions of executing workloads, as explained in Section~\ref{sec:computing_footprint}.
However, close alignment with variable renewable generation will still be beneficial given energy storage losses, so we expect carbon-aware demand management techniques to remain relevant.  

\textbf{Take away}: Variable renewable energy generation, fluctuating grid demand, and the dynamic availability of compute resources should be taken into account to minimise the footprint of scientific workflows in the near term.

\subsection{Economical Impacts and Rebound Effects}

Improving the energy efficiency of tasks and workflows often aligns with economic objectives.
If scientific workflows are executed on a cluster, lower energy consumption by workflows directly translates to lower energy bills.
This is beneficial for organisations, even if individual researchers or research groups do not directly pay for their energy consumption.
If the workflows are executed using public cloud resources, reduced energy bills will benefit cloud providers, but improved energy efficiency will also often coincide with reduced consumer costs when tasks and workflows make better use of paid-for resources and complete faster.

Whether there will also be significant cost savings as a result of lower emissions remains to be seen.
Some companies have ``net zero'' targets and are committed to offsetting any emissions still incurred, including Scope 3 emissions as reported by cloud providers.
The BBC, for example, is targeting to reach net zero emissions by 2030, by first reducing emissions and then offsetting the remainder, explicitly considering cloud services as part of their Scope 3 emissions\footnote{\url{https://www.bbc.co.uk/sustainability/our-plan/}, accessed 2025-05-15}.
In such circumstances, improving the energy efficiency of individual tasks and entire workflows will lower emissions and, hence, decrease the cost of offsetting the remaining emissions.
Furthermore, the effectiveness of offsetting is highly disputed\footnote{see e.g. \url{https://www.source-material.org/vercompanies-carbon-offsetting-claims-inflated-methodologies-flawed/}, accessed 2024-05-10}.
The UKRI funding agency, for example, will not support any offsetting of emissions from research projects, as there is insufficient evidence that they have long-term impacts\footnote{\url{https://www.ukri.org/who-we-are/policies-standards-and-data/corporate-policies-and-standards/environmental-sustainability/ukri-position-statement-on-carbon-offsetting/}, accessed 2024-05-08}.
Reducing emissions in the first place is, therefore, both a less expensive and a safer approach.

Improved efficiencies and reduced costs carry the risk of rebound effects.
Put simply, organisations, research groups, and scientists may spend resources or budgets freed up by efficiency gains immediately on running additional workflows.
In fact, there may even be a net increase in energy consumption and carbon emissions, as improved efficiencies can make use cases viable that were not before.  
This effect is known as \emph{Jevons' paradox}.
It is therefore crucial that scientists remain aware of the significant emissions that remain, even with more energy-efficient workflows.

\textbf{Take away}: Reducing the energy consumption and emissions from workflows often coincides with economic objectives, but we must not forget the urgent need to decarbonise society, so any savings achieved cannot mean we can run more workflows.

\section{Conclusion}
\label{sec:conclusion}

Scientific workflow applications often require substantial compute infrastructure and energy, both of which can be linked to carbon emissions. As described in Section~\ref{sec:problem_analysis}, the operational emissions of a single run of a geoscience workflow on a sample input can be estimated to be around 6.95 kg $CO_{2}e$, whereas the emissions from an astronomy workflow experiment can be calculated to be around 350 kg $CO_{2}e$.
Moreover, we estimated that a bioinformatics task instance can be associated with almost 4.5 times less energy consumption and operational emissions when executed on a public cloud VM in comparison to a ten-year old cluster node.

There are a variety of techniques for reducing the energy consumption and emissions of workflows.
On the one hand there are well-established techniques that improve the energy-efficiency of individual tasks (Section~\ref{sec:individual_task_techniques}), enabling optimised code generation for the available processing platforms, including compiling code for the most energy-efficient compute node within a heterogeneous system.
On the other hand, a wealth of research has investigated energy-aware scheduling, placement, and resource right-sizing for scientific workflows (Section~\ref{sec:entire_workflow_techniques}), commonly leveraging models of the performance and energy efficiency of tasks on different compute resources.

We conclude that there is a clear need to reduce the carbon footprint of scientific workflows running on compute clusters, while there are numerous techniques to achieve this.
Still, to the best of our knowledge, few of these techniques are widely used in practice.
We therefore end our chapter with a call to action:

\begin{itemize}
    \item to care about the environmental impact of scientific workflows,
    \item to research and develop practical techniques to save energy and emissions in scientific compute clusters,
    \item and to employ available techniques in practice to minimise the environmental impact of scientific data analysis.
\end{itemize}

Scientific workflows may not dominate computing workloads on a global scale.
Data analysis workflows may also be vital to the complex research being conducted in many fields.
Nevertheless, as scientists, we must lead by example and push the boundaries of what is possible, not only in our respective fields, but also in the sustainability of our research practices.

\section*{Acknowledgements}

This work was supported by the Engineering and Physical Sciences Research Council under grant number UKRI154.

We also gratefully acknowledge the main sources of energy estimates and carbon intensity data used in Section~\ref{sec:problem_analysis}: \url{https://www.cloudcarbonfootprint.org/} (various estimates) and \url{https://www.electricitymaps.com/} (historical data only). 

Furthermore, we thank Jonathan Bader and Fabian Lehmann for their valuable inputs into the calculations in Section~\ref{sec:problem_analysis}, and Kathleen West for her constructive feedback on drafts of the same section. We would also like to thank the anonymous reviewers for their thoughtful comments.

\printbibliography{myChapter}

\end{document}